\documentstyle [11pt] {article}

\renewcommand{\theequation}{\thesection.\arabic{equation}}

\begin{document}

\title{Gravitational waves propagating into
Friedmann--Robertson--Walker universes}

\author{by \ Ji\v{r}\'\i \ Bi\v{c}\'ak\thanks{E--mail:
bicak@hp03.troja.mff.cuni.cz} \\
\\Department of Theoretical Physics,
Faculty of Mathematics and Physics,\\
Charles University, V Hole\v sovi\v ck\'ach 2, 
18000 Prague 8, Czech Republic\\
\\
and \ Jerry B. Griffiths\thanks{E--mail: J.B.Griffiths@Lboro.ac.uk} \\ \\
Department of Mathematical Sciences,
Loughborough University,\\
Loughborough, Leicestershire. LE11 3TU, U.K. \\}

\maketitle

\bigskip\bigskip
\begin{abstract}
\baselineskip=18pt
We consider space-times with two isometries which represent gravitational
waves with distinct wavefronts which propagate into exact
Friedmann--Robertson--Walker (FRW) universes. The geometry of possible
wavefronts is analysed in detail in all three types of FRW models. In the
spatially flat and open universes, the wavefronts can be planar or
cylindrical; in the closed case they are toroidal. Exact solutions are given
which describe gravitational waves propagating into the FRW universes with a
fluid with a stiff equation of state. It is shown that the plane-fronted
waves may include impulsive or step (shock) components, while the cylindrical
waves in the spatially flat and open universes and the toroidal waves in
closed universes may contain steps. In general, wavefronts may exist which
have an arbitrary finite degree of smoothness. In all cases, the waves are
backscattered. The head-on collision of such waves is also briefly mentioned. 
\end{abstract}

\medskip\vfil\eject

\section{Introduction}
\setcounter{equation}{0}
\baselineskip=15pt

Most work on gravitational waves in cosmological backgrounds has been based
on first order approximation methods. However, if gravitational radiation
played a significant role in the universe, then the waves would cause the
symmetries of the standard homogeneous and isotropic models to be violated.
In order to treat general, large perturbations one has to turn to a numerical
approach. However, when space-time is assumed to have two spacelike Killing
vectors, exact solutions containing gravitational waves or solitons can be
constructed, at least for vacuum or for scalar or electromagnetic fields or
for a stiff perfect fluid in which the pressure is equal to the density. Some
of these solutions have been reviewed for example by Carmeli {\it et al}
\cite{CCM81}, by Adams {\it et al} \cite{AHZ-82}, and recently by Verdaguer
\cite{Ver93}.

In previous work on gravitational waves in cosmological backgrounds, the
exact solutions considered have been interpreted as perturbations extending
over whole universes. As such, it has not been possible to give an analysis
of the geometry of the wave surfaces. In this paper we present the new class
of solutions representing gravitational waves propagating into homogeneous,
isotropic universes. The space-time ahead of a gravitational wave is taken to
be an exact Friedmann--Robertson--Walker (FRW) universe. Since this has
spatial sections of constant curvature, the geometrical properties of the
wavefront can be explicitly determined. For these solutions, we can also
analyse the character of the wavefront: whether it contains a
$\delta$-function impulse, or a shock (step) wavefront, or is smooth in the
sense that the first $n$ derivatives of the Weyl tensor are continuous across
the front.

We consider space-times which admit two commuting, hypersurface orthogonal
spacelike Killing vectors. These can be described by the line element
 \begin{equation}
 \label{1.1} 
  ds^2 =e^{-M}(d\eta^2-d\mu^2) -e^{-U}(e^V{dx}^2+e^{-V}{dy}^2), 
 \end{equation}
 where the metric functions $U$, $V$ and $M$ depend on the coordinates $\eta$
and $\mu$ only. In fact, all the FRW models can also be written in this form
for a perfect fluid with any equation of state, with or without a
cosmological constant. Thus, certain exact solutions described by this line
element can be considered to be (large) perturbations of these FRW models,
provided they reduce to particular models in appropriate limits. (Other
classes of inhomogeneous universes which can reproduce the FRW models in
certain limits have recently been reviewed by Krasi\'nski \cite{Kra95}.)

It may immediately be noticed that, in the form of the line element
(\ref{1.1}), two families of null hypersurfaces, $\eta\pm\mu=$~const., are
singled out in a natural way. These surfaces may be regarded as wave surfaces
for both approximate and exact gravitational waves. The geometry of these
wavefronts in the FRW universes of all three types will be analysed in detail
in this paper. These null hypersurfaces may also be considered as suitable
wavefronts for some other radiative fields.

In a series of recent papers \cite{Gri93a}--\cite{FeGr94}, some families of
exact solutions have been outlined which describe the propagation (and
collision) of gravitational waves into FRW backgrounds in the above form. In
these solutions, the space-time is exactly FRW with stiff fluid ahead of the
waves. Such waves therefore have well-defined wavefronts. It was previously
noted that these waves are necessarily backscattered, and that their
interactions following head-on collisions do not induce additional
singularities into the space-time. However, complete metrics were not given
and the structure of the wavefront was not analysed.  A method for obtaining
complete solutions was later described in \cite{AlGr95}.

It is the purpose of this paper to complete the derivation of the solutions
representing waves with distinct wavefronts propagating into FRW backgrounds,
and to summarise them from a unified point of view for each possible type of
wavefront and all three types of FRW models. We will describe the waves by
considering the most general form of the solution of the main field equation
for a distinct wavefront. Using the method outlined in \cite{AlGr95}, we will
complete the integration of the subsidiary equations, thus determining the
whole metric. We will also analyse the structure of the wavefront in each
case. It will be shown that these may include impulsive components in some
cases, but may contain step (or shock) waves in all cases. Waves in which the
Weyl tensor has $n$ continuous derivatives are also discussed.

We will first describe the geometrical character of possible wavefronts in
each FRW model in order to treat the properties of all the above solutions
from a unified point of view. In Sections 2--4, we will consider the
spatially flat, open and closed FRW models in turn. It will be shown that, in
the spatially flat and open FRW universes, the gravitational waves can have
plane wavefronts. In these universes, the gravitational wavefronts can
alternatively be cylindrical. In the case of the closed FRW universe, the
wavefronts are toroidal. In the Appendix, we describe in detail how a 2-torus,
forming a wavefront, can be constructed in the closed model. The character of
the wavefronts reflects only the symmetries assumed. This approach is thus
relevant also in any study of more general waves and cosmological models than
those treated here. It may be of interest to analyse even small (linear)
perturbations of FRW universes with these symmetries, with other equations of
state and, possibly, with a non-vanishing cosmological constant.

After summarizing the properties of the FRW universes with a stiff fluid in
Section~5, we study in Section~6 the general family of solutions which
describe waves with distinct wavefronts propagating into these backgrounds.
Then, in Sections 7--11, we analyse respectively the separate cases of plane
and cylindrical waves in spatially flat and open universes, and toroidal waves
in a closed universe.

As well as analysing the character of the wavefront, we also find that, for
the waves with plane wavefronts the only space-time singularity is that
describing the big bang. However, our cylindrical and toroidal waves will, in
general, propagate away from (and be backscattered towards) a singular axis,
which can be considered as both ``source'' and ``absorber'' of the waves.
Other remarks appropriate by way of conclusion are collected in Section~12.

\section{Wavefronts in spatially flat FRW universes}
\setcounter{equation}{0}

In the spatially flat case (with curvature index $k=0$), the
Robertson--Walker line element may be written in either of the forms
 \begin{eqnarray}
 ds^2 &=& dt^2-R^2(t)\big(dx^2+dy^2+dz^2\big), \label{2.1a}  \\
 ds^2 &=& dt^2-R^2(t)\big(d\rho^2+\rho^2d\phi^2+dz^2\big),  \label{2.1b} 
 \end{eqnarray} 
 where $R(t)$ is determined by Friedmann's equation. In order to consider
null hypersurfaces, it is convenient to introduce the ``conformal time''
coordinate $\eta$ such that 
 \begin{equation}
 \label{2.2} 
dt=R(\eta)d\eta. 
 \end{equation}
 The line elements (\ref{2.1a}) and (\ref{2.1b}) can then be written in
the form (\ref{1.1}):
\begin{eqnarray}
 ds^2 &=& R^2(\eta)\big(d\eta^2-dz^2\big) -R^2(\eta)\big(dx^2+dy^2\big),
 \label{2.3}  \\
 ds^2 &=& R^2(\eta)\big(d\eta^2-d\rho^2\big)
-R^2(\eta)\big(\rho^2d\phi^2+dz^2\big).  \label{2.4} 
\end{eqnarray} 
 In (\ref{2.3}), surfaces on which $\eta=$~const., $z=$~const. are planes,
and the null hypersurfaces $\eta-z=$~const. represent plane wavefronts.
Clearly, plane wave surfaces oriented in any direction can immediately be
constructed. Alternatively, in the form (\ref{2.4}), surfaces on which
$\eta=$~const., $\rho=$~const. are cylinders, and the null hypersurfaces
$\eta-\rho=$~const. represent expanding cylindrical wavefronts.

\section{Wavefronts in FRW open universes} 
\setcounter{equation}{0}

The standard form of the metric of an open FRW model with negative spatial
curvature $(k=-1)$ in coordinates $\chi\in[0,+\infty)$, $\theta\in[0,\pi)$,
$\phi\in[0,2\pi)$ reads
 \begin{equation}
 \label{3.1} 
ds^2 =dt^2 - R^2(t)\big[d\chi^2 + \sinh^2\chi({\rm
d}\theta^2 +\sin^2\theta\, d\phi^2)\big]. 
 \end{equation}
 As is well known \cite{MTW73}, the spatial 3-geometry on a hypersurface
$t=t_0=$~const.,
 \begin{equation}
 \label{3.2} 
d\ell^2 ={R_0}^2 \big[d\chi^2 + \sinh^2 \chi(d\theta^2 +
\sin^2\theta\, d\phi^2)\big],
\end{equation}
 where $R_0=R(t_0)$, can be represented as 3-dimensional hyperboloid
 \begin{equation}
 \label{3.3}  
V^2 - X^2 - Y^2 - Z^2 = {R_0}^2, 
\end{equation}
 embedded in a 4-dimensional Minkowski space with Lorentzian coordinates
$V,X,Y,Z$, where 
 \begin{eqnarray} 
 \label{3.4} 
V &=& R_0\cosh\chi, \nonumber\\
Z &=& R_0\sinh\chi\cos\theta,\nonumber\\
X &=& R_0\sinh \chi \sin\theta \cos\phi,\nonumber\\
Y &=& R_0\sinh\chi\sin\theta \sin\phi. 
\end{eqnarray}

In order to find 2-planes in the open FRW model, we introduce alternative
coordinates in which the 3-metric is conformally flat. By putting
 \begin{eqnarray}
 \label{3.5} 
 X &=& R_0{x\over z}, \qquad Y =R_0{y\over z},\nonumber\\
Z &=& {R_0\over 2z} (1 - x^2 - y^2 - z^2),\\
V &=& {R_0\over 2z} (1 + x^2 + y^2 + z^2),\nonumber
 \end{eqnarray}
 where $x, y\in(-\infty, \infty)$, $z\in (0, +\infty)$, we find (\ref{3.3}) to
be satisfied. These coordinates sweep out the whole ``positive'' $(V>0)$ sheet
of the hyperboloid (\ref{3.3}). The 3-metric now takes the form
 \begin{equation}
 \label{3.6} 
 d\ell^2 = {{R_0}^2\over z^2}(dx^2 +dy^2+dz^2). 
 \end{equation}
 Clearly, putting $z=$~const. we obtain 2-surfaces with intrinsically flat
geometry. Both the 3-metric (\ref{3.6}) and the whole space-time metric
admit, on these 2-surfaces, not only the translational Killing vectors
$\partial_x$ and $\partial_y$, but also the rotational Killing vector
$y\partial_x-x\partial_y$. The proper lengths along the whole orbits of both
$\partial_x$ and $\partial_y$ are infinite. Thus, these 2-surfaces on which
$z=$~const. can be considered to be 2-planes.

In figure~1 we illustrate these 2-planes in the embedding diagram of the
section $Y=0$, or $\phi=0$ and $\pi$, through the 3-hyperboloid (\ref{3.3})
in a 3-dimensional Minkowski space with coordinates $V,X,Z$. Since
$z=z_0=$~const. implies $V+Z=R_0/z_0=$~const., the 2-planes are formed by
cutting the 3-hyperboloid (\ref{3.3}) along ``null hyperplanes''
$V+Z=$~const. in the Minkowski space. In this way we obtain the 2-planes
described in coordinates $V,X,Y,Z$ by 
 \begin{eqnarray}
 \label{3.7} 
 -Z &=& {\textstyle{1\over2}}z_0(X^2+Y^2)+Z_0, \nonumber\\
Z_0 &=& {\textstyle{1\over2}}z_0{R_0}^2-{z_0}^{-1}={\rm const.}, \\
V &=& R_0{z_0}^{-1}-Z. \nonumber
 \end{eqnarray}
 For a fixed $z_0$, this is a 2-paraboloid which in figure~1 reduces to a
parabola, the $Y$ dimension having been suppressed. (Figure~1 remains the
same after replacing $X$ by $Y$.) On the other hand, by considering sections
of Minkowski space on which $Z=$~const., equations (\ref{3.5}) and
(\ref{3.7}) for fixed $z=z_0$ imply $X^2+Y^2=$~const., or $x^2+y^2=$~const.
With these cuts of the 3-hyperboloid, the infinity of the 2-planes cannot be
seen. Indeed, in terms of the original coordinates $\chi,\theta,\phi$, the
condition $V+Z=R_0{z_0}^{-1}$ reads
$\cosh\chi+\sinh\chi\cos\theta={z_0}^{-1}$. In order to consider
$\chi\to\infty$, we have to let $\theta\to\pi$, as expected for paraboloids
with axes in that direction. In choosing coordinates according to
(\ref{3.5}), we ``singled out'' the coordinate $z$; since the 3-hyperboloid
is isotropic and homogeneous, we can construct other sets of 2-planes by
simple transformations.

Cylindrical 2-surfaces in the open FRW models can easily be found by
introducing coordinates $\rho\in[0,+\infty)$, $y\in(-\infty,+\infty)$,
$\phi\in[0,2\pi)$ such that
 \begin{eqnarray}
 \label{3.8} 
 V &=& R_0\cosh\rho\cosh y, \nonumber\\
Z &=& R_0\cosh\rho\sinh y, \nonumber\\
X &=& R_0\sinh\rho\cos\phi, \nonumber\\
Y &=& R_0\sinh\rho\sin\phi. 
\end{eqnarray} 
 The 3-metric on the 3-hyperboloid then takes the form
 \begin{equation}
 \label{3.9} 
 d\ell^2={R_0}^2 \big[d\rho^2+\sinh^2\rho\,d\phi^2+\cosh^2\rho\,dy^2\big].
 \end{equation}
 Putting $\rho=\rho_0=$~const., we obtain cylindrical 2-surfaces. The proper
length along the orbits of the Killing vector $\partial_\phi$ is finite,
being $2\pi R_0\sinh\rho$, but it is infinite along the orbits of
$\partial_y$. Although the local geometry on the cylinder is intrinsically
flat, there exists no other Killing vector in this 2-space as occurs in the
case of the 2-plane.

In figure~2, a cylindrical 2-surface $\rho=\rho_0$ is illustrated in the same
embedding diagram as in figure~1. From (\ref{3.8}), we now find
 \begin{eqnarray}
 \label{3.10} 
 X^2+Y^2 &=& (R_0\sinh\rho_0)^2={\rm const.}, \nonumber\\
V^2-Z^2 &=& (R_0\cosh\rho_0)^2={\rm const.} 
 \end{eqnarray}
 Since the $Y$-dimension is suppressed ($\phi=0$ and $\pi$) in figure~2, we
obtain $X=\pm R_0\sinh\rho_0$ and $V^2-Z^2=(R_0\cosh\rho_0)^2$. These are
two hyperbolae, representing two generators of the cylindrical 2-surface
$\rho=\rho_0$. The axis of the cylinder is given by $\rho=0$, i.e. $\theta=0$
or $\pi$, or $X=Y=0$. Considering the section of the 3-hyperboloid
(\ref{3.3}) on which $Z=0$, the 2-cylinder would be represented by the circle
$X^2+Y^2=(R_0\sinh\rho_0)^2$, $V=R_0\cosh\rho_0$, as follows from
(\ref{3.10}). Clearly, for $\rho_0$ small, the 2-cylinder would approximate
to an ordinary cylinder in $E^3$.

In terms of the two new sets of coordinates (\ref{3.5}) and (\ref{3.8}), the
line element (\ref{3.1}) takes the following forms 
 \begin{eqnarray}
 ds^2  &=& dt^2-R^2(t) \left[{1\over z^2}(dx^2+dy^2+dz^2)\right],
 \label{3.11}  \\
 ds^2  &=& dt^2-R^2(t)
\left[d\rho^2+\sinh^2\rho\,d\phi^2+\cosh^2\rho\,dy^2\right]. \label{3.12} 
 \end{eqnarray}
 Introducing the new coordinate
 \begin{equation}
 \label{3.13} 
\mu = \log z, 
 \end{equation}
 the line element (\ref{3.11}) becomes
 \begin{equation}
 \label{3.14} 
 ds^2=dt^2 - R^2(t) \left[d\mu^2 + e^{-2\mu}(dx^2+dy^2)\right].
 \end{equation}
 Finally, going over to the ``conformal time'' coordinate $\eta$ in both
cases, we arrive at the metric 
 \begin{equation}
 \label{3.15} 
 ds^2=R^2(\eta) \big(d\eta^2-d\mu^2\big)
-R^2(\eta)e^{-2\mu}\big(dx^2+dy^2\big). 
 \end{equation}
 Alternatively, the line element (\ref{3.12}) can be written as
 \begin{equation}
 \label{3.16} 
 ds^2=R^2(\eta) \big(d\eta^2-d\rho^2\big) 
 -{\textstyle{1\over2}}R^2(\eta)\sinh2\rho\big(\tanh\rho\,d\phi^2
 +{\rm coth}\,\rho\,dy^2\big).
 \end{equation}
 Both metrics (\ref{3.15}) and {\ref{3.16}) are now expressed in the form
(\ref{1.1}).

\section{Wavefronts in FRW closed universes} 
\setcounter{equation}{0}

We start with the standard form of the metric for a closed FRW model
having positive spatial curvature ($k=1$)
 \begin{equation}
 \label{4.1} 
  ds^2=dt^2-R^2(t) \big[d\chi^2+\sin^2\chi
\big(d\theta^2+\sin^2\theta\,d\phi^2\big)\big], 
 \end{equation}
 where $\chi\in[0,\pi]$, $\theta\in[0,\pi]$, $\phi\in[0,2\pi]$. It is well
known that the spatial 3-geometry on a hypersurface of homogeneity $t=t_0=$
const., 
 \begin{equation}
 \label{4.2} 
  d\ell^2=R_0^2 \big[d\chi^2+\sin^2\chi
\big(d\theta^2+\sin^2\theta\,d\phi^2\big)\big], 
 \end{equation}
 where $R_0=R(t_0)$, can be visualised as a 3-dimensional sphere
 \begin{equation}
 \label{4.3} 
  W^2+X^2+Y^2+Z^2=R_0^2, 
 \end{equation}
 embedded in a 4-dimensional Euclidean space with cartesian coordinates $W$,
$X$, $Y$, $Z$. By setting
 \begin{eqnarray}
 \label{4.4} 
 W &=& R_0\cos\chi, \nonumber\\
Z &=& R_0\sin\chi\cos\theta, \nonumber\\
X &=& R_0\sin\chi\sin\theta\cos\phi, \nonumber\\
Y &=& R_0\sin\chi\sin\theta\sin\phi, 
\end{eqnarray}  
 Eq. (\ref{4.3}) is satisfied.

In the Appendix it is demonstrated in detail that, with the symmetries
assumed, a typical wave surface in a closed FRW model is a 2-torus in the
3-sphere. This is best described by parametrizing the whole 3-sphere by
coordinates $\zeta\in[0,\pi/2]$, $\sigma\in[0,2\pi]$, $\delta\in[0,2\pi]$
such that (cf. Eq. (\ref{A.5}) in the Appendix)
 \begin{eqnarray}
 \label{4.5} 
 W &=& R_0\cos\zeta\cos\sigma, \nonumber\\
Z &=& R_0\cos\zeta\sin\sigma, \nonumber\\
X &=& R_0\sin\zeta\cos\delta, \nonumber\\
Y &=& R_0\sin\zeta\sin\delta. 
\end{eqnarray} 
 The 2-torus is specified by the choice of the parameter $\kappa=\cos\zeta$.
(For an illustration, see Figure~8 in the Appendix.)

In terms of the new coordinates, the line element (\ref{4.1}) takes the form
 \begin{equation}
 \label{4.6} 
 ds^2=dt^2-R^2(t) \big[d\zeta^2 +\sin^2\zeta\,d\delta^2
+\cos^2\zeta\,d\sigma^2\big], 
 \end{equation}
 or, using the conformal time $\eta$,
 \begin{equation}
 \label{4.7} 
 ds^2 = R^2(\eta)\big(d\eta^2-d\zeta^2\big) 
-{\textstyle{1\over2}}R^2(\eta)\sin2\zeta \big(\tan\zeta\,d\delta^2
+{\rm cot}\,\zeta\,d\sigma^2\big), 
 \end{equation}
 which is clearly in the form (\ref{1.1}).

\section{FRW universes with a stiff fluid} 
\setcounter{equation}{0}

We now consider the case of a ``stiff'' perfect fluid in which the pressure
is equal to the density. This equation of state may be considered as a
``modern version of aether'' (see \cite{Kra95}) which enables mathematical
relativists to generate a variety of new solutions. However, it may also have
played an important role in the period of high densities after the big bang
\cite{Zel62}, \cite{Bar78}, or even in the interior of cold condensed
objects. Its role as a limiting equation of state of neutron matter was
studied thoroughly in the full field theory by Walecka \cite{Wal74}.
Difficult and important problems arise, of course, when studying
gravitational waves which propagate in dust or in fluids with a more
realistic equation of state. However, in the case of a stiff fluid, families
of exact radiative solutions can be obtained which can be considered to be
perturbations of the FRW models with a stiff fluid. In this section, we
review the unperturbed background models.

The equation of state of the fluid is taken to be $\tilde p=\tilde\rho$. (We
use tildes to denote quantities referring to the fluid to avoid confusion
with the same symbols used elsewhere.) With this, the equation of energy
conservation $(\tilde\rho R^3),_t=-\tilde p(R^3),_t$ implies that
 \begin{equation}
 \label{5.1} 
 \tilde\rho R^6 =\tilde\rho_0R_0^6 ={\rm const.} =\gamma_0
={\textstyle{3\over8\pi}}\gamma^2,
 \end{equation}
 where the dimension of the constant $\gamma$ is (length)$^2$. The standard
Friedmann equation, ${R,_t}^2=-k+\gamma^2R^{-4}$, $k=0,\pm1$, can then easily
be integrated after introducing the conformal time coordinate $\eta$ in
accordance with (\ref{2.2}). The solutions read \cite{Bel79}, \cite{BiGr94}
 \begin{eqnarray}
 \label{5.2} 
R &=& [\gamma\sinh 2\eta]^{1\over 2},\quad \hbox{if \ $k = -1$}, \nonumber\\
R &=& [\gamma2\eta]^{1\over 2},\qquad\quad \hbox{if \ $k = 0$}, \\
R &=& [\gamma\sin 2\eta]^{1\over 2},\quad \ \hbox{if \ $k = +1$}. \nonumber 
 \end{eqnarray}

In \cite{BiGr94} the physical time $t$ is expressed explicitly in terms of
$\eta$ by means of elliptic integrals in the cases $k=\pm1$. For the
spatially flat case one gets simply
 \begin{equation}
 \label{5.3} 
 R = (3\gamma t)^{1\over 3}, 
 \end{equation}
 so that the flat FRW universe with a stiff fluid expands slower than the
corresponding universe with dust for which $R\sim t^{2\over 3}$ (see e.g.
\cite{MTW73}). For the open universe, one finds \cite{BiGr94} $R\cong t$ at
large times, i.e. exactly the same asymptote as the standard open Friedmann
universes with dust or radiation. All these open universes thus approach the
flat Milne universe, for which $R=t$ exactly.

The ``gravitational role'' of the pressure in slowing down the expansion in
the spatially flat case can also be seen in the ``speeding up'' of the
recollapse in the case of the closed FRW models. The conformal time $\eta$ is
also often called the ``arc parameter'' since it determines the radius of arc
distance on the 3-sphere covered by a photon travelling since the start of
the expansion \cite{MTW73}. In the case of the closed universe with dust, the
range of $\eta$ is $2\pi$ --- a photon makes ``one trip'' around the whole
universe before its final collapse. In the model with radiation, in which
$\tilde p=\tilde\rho/3$, the range of $\eta$ is $\pi$ --- a photon only gets
as far as the antipodal point of the universe. In the universe with a stiff
fluid, the last equation in (\ref{5.2}) shows that the range of $\eta$ is
$\pi/2$ --- a photon succeeds in traveling only one quarter of the distance
around the whole universe (say, from the pole to the equator). This will be
relevant in Section~11 where we shall consider exact toroidal waves
propagating into a closed FRW background with a stiff fluid.

It is easy to write the line element of all three types of FRW models in
the form (\ref{1.1}). Since $\gamma$ is an arbitrary constant with dimension
(length)$^2$, we may consider (\ref{1.1}) in the dimensionless form
$ds^2/\gamma^2$. Without loss of generality we may thus put $\gamma=1$.
Comparing now (\ref{2.3}), (\ref{2.4}), (\ref{3.15}), (\ref{3.16}) and
(\ref{4.7}) with $R(\eta)$ substituted from (\ref{5.2}) with the general form
(\ref{1.1}), we find the following:

\medskip\noindent {\bf Case 1a.} \ 
In the spatially flat case with the plane 2-surfaces $z=$~const.,
we have $\mu=z\in(-\infty,+\infty)$, $(x,y)\in{\hbox{$I\kern-3pt R$}}^2$,
$\eta\in[0,\infty)$, and
 \begin{equation}
 \label{5.4} 
  e^{-U}=2\eta, \qquad e^{-M}= 2\eta, \qquad V=0. 
\end{equation}

\medskip\noindent {\bf Case 1b.} \ 
In the spatially flat case with the cylindrical 2-surfaces $\rho=$~const.,
$\mu=\rho\in[0,\infty)$,
$x=\phi\in[0,2\pi)$, $y=z\in(-\infty,+\infty)$, $\eta\in[0,\infty)$, and
 \begin{equation}
 \label{5.5} 
 e^{-U}=2\eta\rho, \qquad e^{-M}= 2\eta, \qquad e^V=\rho. 
 \end{equation}

\medskip\noindent {\bf Case 2a.} \ 
In the open model with the plane 2-surfaces $z=$~const. or $\mu=$~const., we
have $\mu\in(-\infty,+\infty)$, $(x,y)\in{\hbox{$I\kern-3pt R$}}^2$,
$\eta\in[0,\infty)$, and
 \begin{equation}
 \label{5.6} 
 e^{-U}=\sinh2\eta\,e^{-2\mu}, \quad e^{-M}=\sinh2\eta, \quad V=0. 
\end{equation}

\medskip\noindent {\bf Case 2b.} \ 
In the open model with the cylindrical 2-surfaces $\rho=$~const.,
$\mu=\rho\in[0,\infty)$,
$x=\phi\in[0,2\pi)$, $y=z\in(-\infty,+\infty)$, $\eta\in[0,\infty)$, and
 \begin{equation}
 \label{5.7} 
 e^{-U}=\sinh2\eta\sinh2\rho, \quad e^{-M}=\sinh2\eta, \quad e^V=\tanh\rho.
 \end{equation}

\medskip\noindent {\bf Case 3.} \ 
In the closed model with the toroidal 2-surfaces
$\zeta=$~const., we have $\mu=\zeta\in[0,\pi/2]$,
$x=\delta\in[0,2\pi)$, $y=\sigma\in[0,2\pi)$, $\eta\in[0,\pi/2]$, and
 \begin{equation}
 \label{5.8} 
 e^{-U}=\sin2\eta\sin2\zeta, \quad e^{-M}=\sin2\eta, \quad
e^V=\tan\zeta. 
 \end{equation}
\medskip

It is well known \cite{WIM79} that, in a space-time with the metric
(\ref{1.1}), a stiff perfect fluid can be associated with a scalar potential
$\tilde\sigma(\eta,\mu)$ such that the density and 4-velocity of the fluid
are given by 
 \begin{equation}
 \label{5.9} 
 16\pi\tilde\rho=e^M({\tilde\sigma_\eta}^2-{\tilde\sigma_\mu}^2), \qquad
\tilde u_\alpha
=\tilde\sigma_{,\alpha} /(\tilde\sigma_{,\beta}\tilde\sigma^{,\beta})^{1/2},
 \end{equation}
 and, as a consequence of Einstein's equations, the fluid potential
$\tilde\sigma$ satisfies
 \begin{equation}
 \label{5.10} 
 \tilde\sigma_{\eta\eta} -U_\eta\tilde\sigma_\eta -\tilde\sigma_{\mu\mu}
+U_\mu\tilde\sigma_\mu =0.
 \end{equation}

For the above FRW universes with a stiff fluid, the expressions for
$\tilde\sigma$ are found to be 
 \begin{eqnarray}
\tilde\sigma =3^{1/2}\log\eta  \qquad\qquad &&{\rm for }\quad k=0,
\label{5.11}\\
\tilde\sigma =3^{1/2}\log\tanh\eta \qquad &&{\rm for }\quad k=-1, 
\label{5.12} \\
\tilde\sigma =3^{1/2}\log\tan\eta \ \ \qquad &&{\rm for }\quad k=+1. 
 \label{5.13} 
 \end{eqnarray}
 In the spatially flat and open cases, it should be noted that these
expressions for $\tilde\sigma$ apply both in the plane and cylindrical cases.
Clearly, all coordinate systems above are comoving with the fluid.

 Using these background metric functions, we now turn to the construction of
the waves which propagate into these backgrounds.

\section{Waves in FRW universes with a stiff fluid} 
\setcounter{equation}{0}

We return to the general line-element (\ref{1.1}). In the case of a stiff
perfect fluid, and vanishing cosmological constant, Einstein's field
equations imply that $e^{-U}$ satisfies the wave equation
\begin{equation}
 \label{6.1} 
  (e^{-U})_{\eta\eta}-(e^{-U})_{\mu\mu}=0, 
 \end{equation}
 and $V$ satisfies the linear equation
\begin{equation}
 \label{6.2} 
  V_{\eta\eta}-U_\eta V_\eta -V_{\mu\mu}+U_\mu V_\mu=0. 
 \end{equation}
 It may immediately be noted that, in general, Eq. (\ref{6.1}) can be
integrated to give
 \begin{equation}
 \label{6.3} 
  e^{-U}=f(\eta-\mu)+g(\eta+\mu), 
 \end{equation}
 where $f$ and $g$ are arbitrary functions. In many cases, it is convenient
to adopt $f$ and $g$ as coordinates, at least in some region of space-time in
which they are not constant. The arbitrariness thus retained can be used to
describe families of exact solutions.  It can be seen, however, that
singularities of some type will occur when $f+g=0$. In addition, in a closed
universe, $f$ and $g$ cannot be monotonic functions over the entire
space-time, and it is then more convenient to revert to coordinates that are
more directly related to $\eta$ and~$\mu$.

Taking $f$ and $g$ as coordinates, equations (\ref{6.2}) and (\ref{5.10})
take the form of Euler--Poisson--Darboux equations with non-integer
coefficients
 \begin{eqnarray}
 && (f+g)V_{fg} +{\textstyle{1\over2}}V_f +{\textstyle{1\over2}}V_g=0, 
 \label{6.4} \\
 &&
(f+g)\tilde\sigma_{fg} +{\textstyle{1\over2}}\tilde\sigma_f
+{\textstyle{1\over2}}\tilde\sigma_g=0. \label{6.5} 
 \end{eqnarray}
 For any solutions of these equations, the remaining metric function $M$ can
be found by quadratures. Putting 
 \begin{equation}
  e^{-M}={{f'g'}\over(f+g)^{1/2}}\ e^{-S}, 
 \label{6.6}
 \end{equation}
 the remaining field equations become 
 \begin{equation}
  S_f=-{\textstyle {1\over 2}}(f+g)\big({V_f}^2+{\tilde\sigma_f}^2\big),
\qquad S_g=-{\textstyle {1\over 2}}(f+g)\big({V_g}^2+{\tilde\sigma_g}^2\big), 
 \label{6.7}
 \end{equation}
 which are automatically integrable in view of (\ref{6.4}) and (\ref{6.5}).
However, these equations are also significant in determining conditions on
the derivatives of $V$ and $\sigma$ when considering junction conditions
across possible wavefronts. In addition, care has to be taken on any
hypersurface on which $f'$ or $g'$ are zero. Using a null tetrad system that
is naturally adapted to the coordinates (with the null vectors given by
$\ell_\alpha=2^{-1/2}e^{-M/2}(\eta,_\alpha-\mu_\alpha)$ and
$n_\alpha=2^{-1/2}e^{-M/2}(\eta,_\alpha+\mu_\alpha)$, the non-zero
components of the Weyl tensor are given in terms of the coordinates $f$ and
$g$ by 
  \begin{eqnarray}
 &&\Psi_0 = -{\textstyle{1\over2}} {g'}^2 e^M \left(2V_{gg}
+{3\over(f+g)}V_g -(f+g)V_g\big({V_g}^2+{\tilde\sigma_g}^2\big)\right), 
\nonumber\\
 &&\Psi_2 = -{\textstyle{1\over6}} f'g' e^M \left({3\over(f+g)^2}
-3V_fV_g -\tilde\sigma_f\tilde\sigma_g\right), \\
 &&\Psi_4 = -{\textstyle{1\over2}} {f'}^2 e^M \left(2V_{ff} 
+{3\over(f+g)}V_f -(f+g)V_f\big({V_f}^2+{\tilde\sigma_f}^2\big)\right). 
\nonumber
 \label{6.8}
 \end{eqnarray}

In many cases of interest, it is possible to choose $f$ such that $f=0$
represents a null hypersurface. On such a hypersurface, $\eta-\mu=$~const. It
is then convenient to adopt a gauge such that $e^{-U}=f+g$ takes the same
form on either side of $f=0$, thus facilitating the possible choice of $f$
and $g$ as coordinates. It may then be noticed that, since (\ref{6.4}) is
linear in $V$, we may superpose a background solution of (\ref{6.4}) on
either side of the hypersurface with another solution which is non-zero on
one side only. In this way, we can construct an exact solution which contains
a distinct wavefront which is given by the null hypersurface $f=0$. Of
course, it is necessary to satisfy the appropriate (O'Brien--Synge) junction
conditions \cite{ObSy52} across the wavefront. In this case, these conditions
require that both $V$ and $M$ be continuous across $f=0$. In the cases we are
considering in this paper, it is found using Eqs. (\ref{6.7}) that the
continuity of $M$ implies that $V_f$ must also be bounded on $f=0$.

In the class of solutions we wish to consider, the background is taken in
turns to be appropriate forms of the various FRW models with a stiff fluid
given by (\ref{5.4})--(\ref{5.8}). The additional component of $V$ can then
be interpreted as introducing a purely gravitational wave. In addition, the
expression for $\tilde\sigma$ is taken to be unchanged across the wavefront.
This effectively eliminates the introduction of acoustic waves in the fluid
of the type considered by Tabensky and Taub \cite{TaTa73}. However, since Eq.
(\ref{6.5}) for $\tilde\sigma$ is identical to Eq. (\ref{6.4}) for~$V$,
acoustic waves could also be considered by including the same kind of
additional terms in $\tilde\sigma$. Of course, this equivalence occurs only
for a stiff fluid in which the characteristics for both gravitational and
acoustic waves are the same.

As introduced in \cite{AlGr95}, we now consider the appropriate class of
self-similar solutions of (\ref{6.4}) of the form
 \begin{equation}
 V_k(f,g) =(f+g)^k H_k\left({\textstyle {g-f\over f+g} }\right),
 \label{6.9}
 \end{equation}
 where $k$ is an arbitrary real parameter whose range will be determined
below, and the function $H_k(\zeta)$ is a solution of the equation 
 \begin{equation}
  (1-\zeta^2) H_k''+(2k-1)\zeta H_k'-k^2 H_k = 0.  
 \label{6.10}
 \end{equation}
 This equation admits a class of solutions which can be expressed in
terms of hypergeometric functions $F\big(a,b\,;c\,;z\big)$ such that the
solutions (\ref{6.9}) of (\ref{6.4}) can be expressed in either of the forms 
 \begin{eqnarray}
 (f+g)^k H_k\left({\textstyle {g-f\over f+g} }\right) 
&=& c_k\,{f^{k+1/2}\over{g^{1/2}}}\,
F\left({\textstyle{1\over2}},1+k\,;
{\textstyle{3\over2}}+k\,;-{f\over{g}}\right), \label{6.11} \\
 &=& c_k\,{f^{k+1/2}\over(f+g)^{1/2}}\,
F\left({\textstyle{1\over2}},{\textstyle{1\over2}}\,;
{\textstyle{3\over2}}+k\,;{f\over{f+g}}\right), 
 \label{6.12}
 \end{eqnarray}
 where $c_k={\Gamma(1/2)\over\Gamma(k+3/2)}$. Both forms of this expression
(\ref{6.11}) and (\ref{6.12}) will be useful below, depending on the
appropriate ranges of $f$ and $g$. For integer values of $k$, these functions
take particularly simple forms in which successive functions $H_k(\zeta)$ can
be obtained recursively using the relation
 \begin{equation}
 \label{6.13} 
 H_k(\zeta) =\int_1^\zeta H_{k-1}(\zeta')\,d\zeta', 
 \end{equation}
 with the initial solution $H_0(\zeta)=\cos^{-1}\zeta$ if $f\ge0$, or
$H_0(\zeta)=\cosh^{-1}\zeta$ if $f\le0$. It may be noted that the scaling
constants $c_k$ have been included in (\ref{6.11}) and (\ref{6.12}) to ensure
that no additional scale factor appears in (\ref{6.13}), and we may generally
put $H_k'(\zeta)=H_{k-1}(\zeta)$.

As can be seen from (\ref{6.11}) or (\ref{6.12}), the significant feature of
the solution (\ref{6.9}) is that $V=0$ and $V_f$ is bounded on the null
hypersurface $f=0$ provided $k\ge{1\over2}$. It may thus be matched across
the null hypersurface $f=0$ to the solution $V=0$ on the other side. It then
represents some kind of strong gravitational wave with wavefront $f=0$. As
mentioned above, since (\ref{6.4}) is linear, this solution may be added to
some other background solution to represent the propagation of a
gravitational wave in that background. For FRW stiff fluid backgrounds, the
condition $k\ge{1\over2}$ ensures that both $V$ and $M$ are continuous across
the null hypersurface $f=0$, thus satisfying the O'Brien--Synge junction
conditions~\cite{ObSy52}.

In a series of recent papers \cite{Gri93a}--\cite{AlGr95}, \cite{AlGr96a},
\cite{AlGr96b}, families of exact solutions have been considered which
describe the propagation of gravitational waves in various backgrounds,
together with the head-on collision of such waves and their subsequent
interaction. These solutions may here be re-formulated using solutions of
(\ref{6.4}) in the form
 \begin{equation}
 V(f,g) =V_0(f,g) +\Theta(\eta-\mu) \int_{1/2}^\infty \phi(k) (f+g)^k 
H_k\left({\textstyle {g-f\over f+g} }\right) dk, 
 \label{6.14}
 \end{equation}
 where $V_0(f,g)$ is a solution of (\ref{6.4}) representing some particular
background space-time, $\Theta(\eta-\mu)$ is the Heaviside step function in
which the wavefront $f=0$ is assumed to be given by $\eta-\mu=0$, and
$\phi(k)$ is an arbitrary function. The function $\phi(k)$ may be regarded as
the ``spectral amplitude'' of some arbitrary wave profile. It is subject only
to the condition that the above integral exists.

Let us now turn to the specific cases in which the backgrounds are the
various FRW models with a stiff perfect fluid source. In these cases, the
background regions of course are conformally flat. However, the wave regions
are algebraically general indicating that the propagating gravitational wave
is necessarily backscattered.

\section{Plane waves in a spatially flat FRW model} 
\setcounter{equation}{0}

In terms of the metric (\ref{1.1}), the spatially flat FRW universe with a
stiff fluid is given by (\ref{5.4}) and (\ref{5.11}). Thus $e^{-U}=2\eta$, and
we can put 
 \begin{equation}
\label{7.1}
f=\eta-z, \qquad g=\eta+z. 
 \end{equation}
 We may thus consider a plane wavefront given by $\eta=z$ as described in
Section~2, with the region $\eta<z$ being an exact FRW background into which
a gravitational wave propagates.

Since $V=0$ in the background region in this case, the gravitational wave can
be introduced by putting 
 \begin{equation}
 \label{7.2} 
 V(\eta,z) =\Theta(\eta-z) \int_{1/2}^\infty \phi(k) \eta^k 
H_k\left({\textstyle {z\over\eta} }\right) dk, 
 \end{equation} 
 in which a factor of $2^k$ has been absorbed into the arbitrary function
$\phi(k)$, and $H_k(z/\eta)$ is given by (\ref{6.12}) and its extension.
Using the method described in \cite{AlGr95} and \cite{AlGr96b}, the complete
integral for the metric function $M$ can be expressed as 
 \begin{equation}
 \label{7.3} 
 M(\eta,z) =-\log2\eta +\tilde M\,\Theta(\eta-z), 
 \end{equation} 
 where 
 \begin{equation}
 \label{7.4} 
 \tilde M =-{\textstyle{1\over2}} \int_1^\infty {\eta^n\over n}dn
\int_{1/2}^{n-1/2} \phi(k)\phi(n-k) \left[k(n-k)H_kH_{n-k}
+(1-{\textstyle{z^2\over\eta^2}})H_{k-1}H_{n-k-1}\right] dk, 
 \end{equation} 
 in which the argument of $H_k(z/\eta)$ has been omitted.

The non-zero components of the Weyl tensor in this case are given by 
 \begin{eqnarray}
 \label{7.5} 
 &&\Psi_0 =-{1\over8\eta}e^{\tilde M} \big[V_{\eta\eta}+2V_{\eta z}+V_{zz}
-{\eta\over2}(V_\eta+V_z)^3\big], \nonumber\\
 &&\Psi_2 ={1\over16\eta}e^{\tilde M} \big[{V_\eta}^2-{V_z}^2\big], \\
 &&\Psi_4 =-{1\over8\eta}e^{\tilde M} \big[V_{\eta\eta}-2V_{\eta z}+V_{zz}
-{\eta\over2}(V_\eta-V_z)^3\big]. \nonumber
 \end{eqnarray}
 These are zero in the background region. Near the wavefront where $\eta-z$ is
small, their behaviour is governed by the lowest non-zero term in $k$ in
(\ref{7.2}). Taking this lowest term in the form
 \begin{eqnarray}
 \label{7.6} 
 V(\eta,z) &=& {a_k\over c_k}\, 2^k\eta^k H_k\left({\textstyle {z\over\eta}
}\right) \Theta(\eta-z) \nonumber\\
 &=& a_k\, {(\eta-z)^{k+1/2}\over(2\eta)^{1/2}}\,
F\left({\textstyle{1\over2}},{\textstyle{1\over2}}\,;
{\textstyle{3\over2}}+k\,;{\textstyle{1\over2}(1-{z\over\eta})}\right) 
\,\Theta(\eta-z),
 \end{eqnarray} 
 where $a_k$ is a constant, it can be seen that the Weyl tensor components
behave near the wavefront as 
 \begin{eqnarray}
 \label{7.7} 
 &&\Psi_4 \sim-{a_k(k+{1\over2})\over(2\eta)^{3/2}}\,
(\eta-z)^{k-1/2} \,\delta(\eta-z)
-{a_k(k^2-{1\over4})\over(2\eta)^{3/2}}\, (\eta-z)^{k-3/2}
\,\Theta(\eta-z), \nonumber\\
 &&\Psi_2 \sim -{{a_k}^2(k+{1\over2})\over32\eta^3} (\eta-z)^{2k}
\,\Theta(\eta-z), \\
 &&\Psi_0 \sim-{3a_k\over4(2\eta)^{7/2}} (\eta-z)^{k+1/2}
\,\Theta(\eta-z), \nonumber
 \end{eqnarray}
 where $\delta(\eta-z)$ is the Dirac delta function. This immediately
identifies $\Psi_4$ as the component representing the propagating wave with
wavefront $\eta=z$, $\Psi_0$ is the backscattered wave component propagating
in the opposite spatial direction, while $\Psi_2$ represents the effective
interaction between these two waves. It can also be seen that an impulsive
gravitational wave is included if the lowest term in the expansion for $V$ has
$k={1\over2}$. It has a step (or shock) gravitational wave if the lowest term
has $k={3\over2}$. In general, the Weyl tensor is $C^n$ across the wavefront
if $k_{\rm min}=n+{5\over2}$.

It can be shown that this solution is regular everywhere except at the
initial ``big bang'' singularity $\eta=0$, where both the waves and the
background may be considered to originate.

These solutions were first considered in \cite{Gri93a} for the case in which
$k$ takes half integer values only. However, the complete solution was not
obtained and the possible existence of an impulsive component was not
recognised. The complete solution for a sum of terms with integer values of
$k$ only was given in \cite{AlGr95}. The head-on collision of such waves was
discussed in \cite{Gri93b} using half integer values of $k$ only, but an
explicit expression for the function $M$ in the interaction region cannot be
obtained by the methods considered so far.

\section{Cylindrical waves in a spatially flat FRW model} 
\setcounter{equation}{0}

The above method can also be used to construct cylindrical gravitational
waves in a spatially flat FRW universe with a stiff fluid by expressing it
using the coordinates defined in case~1b of Section~5. In this case, the
metric functions are given in the form (\ref{5.5}) and the fluid by
(\ref{5.11}). Since $e^{-U}=2\eta\rho$ in this case, we can put 
 \begin{equation}
\label{8.1}
f=-{\textstyle{1\over2}}(\eta-\rho)^2, \qquad 
g={\textstyle{1\over2}}(\eta+\rho)^2. 
 \end{equation}

We may now consider a gravitational wave with a cylindrical wavefront given
by $\eta=\rho$ as described in Section~2. Accordingly, the region $\eta<\rho$
is taken to be the exact FRW background into which the wave propagates. Such
a wave may be considered to originate on a line $\rho=0$ at the initial
singularity $\eta=0$. In this case $V=\log\rho$ in the background region, and
a gravitational wave in the general form (\ref{6.14}) can be introduced by
putting 
 \begin{equation}
 \label{8.2} 
 V(\eta,\rho) =\log\rho +\Theta(\eta-\rho) \int_{1/2}^\infty \phi(k)
\eta^k\rho^k  H_k\left({\textstyle {\eta^2+\rho^2\over2\eta\rho} }\right) dk, 
 \end{equation} 
 in which a factor of $2^k$ has again been absorbed into the arbitrary
function $\phi(k)$. (This function, representing a spectral amplitude,
should not be confused with the cylindrical coordinate being used in this
case.) The functions $H_k\big({\eta^2+\rho^2\over2\eta\rho}\big)$ are most
conveniently given here by (\ref{6.11}). Putting
 \begin{equation}
\label{8.3}
V=\log\rho +\tilde V\,\Theta(\eta-\rho), \qquad 
M=-\log2\eta +\tilde M\,\Theta(\eta-\rho), 
 \end{equation}
 it can be shown that the subsidiary equations (\ref{6.7}) can be fully
integrated to yield 
 \begin{eqnarray}
 \label{8.4} 
 &&\hskip-12pt \tilde M =-{\textstyle{1\over2}} \int_{1/2}^\infty \phi(k)
\eta^k\rho^k \left[H_k +{\textstyle{(\eta^2-\rho^2)\over2\eta\rho}
{1\over k}}H_{k-1} \right] dk \nonumber\\
 &&\qquad -{\textstyle{1\over2}} \int_1^\infty {\eta^n\rho^n\over n}dn
\int_{1/2}^{n-1/2} \phi(k)\phi(n-k) \nonumber\\
 &&\qquad\qquad\qquad\qquad\qquad 
\left[k(n-k)H_kH_{n-k}
-{\textstyle{(\eta^2-\rho^2)^2\over4\eta^2\rho^2}} H_{k-1}H_{n-k-1}\right] dk, 
 \end{eqnarray} 
 in which the argument of $H_k$ is taken to be
${\eta^2+\rho^2\over2\eta\rho}$.

In this case, the non-zero components of the Weyl tensor in the region
$\eta\ge\rho$ are given by 
 \begin{eqnarray}
 \label{8.5} 
 &&\hskip-12pt \Psi_0 = -{\textstyle{1\over8\eta}}e^{\tilde M}
\left[\tilde V_{\eta\eta}+2\tilde V_{\eta\rho}+\tilde V_{\rho\rho} 
+{\textstyle{1\over\eta+\rho}}(\tilde V_\eta+\tilde V_\rho) 
-{\textstyle{3\eta\over2(\eta+\rho)}}(\tilde V_\eta+\tilde V_\rho)^2
-{\textstyle{\eta\rho\over2(\eta+\rho)}}(\tilde V_\eta+\tilde V_\rho)^3
\right], \nonumber\\
 &&\hskip-12pt \Psi_2 = -{\textstyle{1\over16\eta}}e^{\tilde M} \left[{\tilde
V_\eta}^2-{\tilde V_\rho}^2 -{\textstyle{2\over\rho}}\tilde V_\rho\right], \\
 &&\hskip-12pt \Psi_4 = -{\textstyle{1\over8\eta}}e^{\tilde M} 
\left[\tilde V_{\eta\eta}-2\tilde V_{\eta\rho}+\tilde V_{\rho\rho} 
+{\textstyle{1\over\eta-\rho}}(\tilde V_\eta-\tilde V_\rho)
-{\textstyle{3\eta\over2(\eta-\rho)}}(\tilde V_\eta-\tilde V_\rho)^2
+{\textstyle{\eta\rho\over2(\eta-\rho)}}(\tilde V_\eta-\tilde V_\rho)^3
\right]. \nonumber
 \end{eqnarray}
 The term $-{1\over8\eta}e^{\tilde M} (\tilde V_\eta-\tilde V_\rho)
\delta(\eta-\rho)$ which would normally appear in the expression for $\Psi_4$
has been omitted because, in this case, $\tilde V_\eta=\tilde V_\rho$ on the
wavefront $\eta=\rho$. It follows that this class of solutions does not
include impulsive cylindrical waves.

Near the wavefront where $\eta-\rho$ is small, the character of the
gravitational wave is determined by the lowest non-zero term in $k$ in
(\ref{8.2}). Taking this lowest term in the form
 \begin{eqnarray}
 \label{8.6} 
 \tilde V(\eta,z) &=& {a_k\over c_k}\, 2^{2k}\eta^k\rho^k  
H_k\left({\textstyle{\eta^2+\rho^2\over2\eta\rho} }\right)  \nonumber\\
 &=& a_k\, {(\eta-\rho)^{1+2k}\over(\eta+\rho)}\,
F\left({\textstyle{1\over2}},1+k\,; {\textstyle{3\over2}}+k\,;
\left({\textstyle{\eta-\rho\over\eta+\rho}}\right)^2\right),
 \end{eqnarray} 
 it can be seen that, near the wavefront, the Weyl tensor components behave as 
 \begin{eqnarray}
 \label{8.7} 
 &&\Psi_4 \sim
-{\textstyle{1\over8}}(2k+1)(4k+1)\,a_k\,\eta^{-2}\,(\eta-\rho)^{2k-1}
\,\Theta(\eta-\rho), \nonumber\\
 &&\Psi_2 \sim -{\textstyle{1\over16}}(2k+1)\,a_k\,\eta^{-3}\,(\eta-\rho)^{2k}
\,\Theta(\eta-\rho), \\
 &&\Psi_0 \sim -{\textstyle{3\over32}}\,a_k\,\eta^{-4}\,(\eta-\rho)^{2k+1}
\,\Theta(\eta-\rho). \nonumber
 \end{eqnarray}
 It can thus be seen that these solutions include a step (or shock)
gravitational wave if the lowest term in the expansion for $V$ has
$k={1\over2}$. Generally, the Weyl tensor is $C^n$ across the wavefront if
$k_{\rm min}={1\over2}n+1$.

It can be shown that $\tilde V$ and $\tilde M$ and their derivatives with
respect to $\eta$ and $\rho$ are all bounded as $\rho\to0$ for all $\eta>0$.
It thus follows from (\ref{8.5}) that the component $\Psi_2$ of the Weyl
tensor is singular on the axis $\rho=0$. Thus the axis of the cylindrical
waves is a scalar curvature singularity even though the fluid density on it
given by (\ref{5.11}) is bounded. This axis may be considered both as the
source of the cylindrical waves, and also as a ``sink'' of the backscattered
radiation.

The possibility of obtaining exact solutions of this type was mentioned in
\cite{AlGr95} but the explicit solutions are given here for the first time.
By superposing two sets of terms similar to those included in (\ref{8.2}) but
with wavefronts given by $\eta=\rho-a$ and $\eta=b-\rho$ for positive
constants $a$ and $b$, it is also possible to consider the collision of
outgoing and incoming cylindrical gravitational waves. This is qualitatively
described in figure~3. Exact solutions describing this situation are only
singular on an initial hypersurface and on the axis of symmetry. However,
although the qualitative features of such a solution can be determined, the
complete integral for $M$ cannot be obtained in this case using the methods
described above.

\section{Plane waves in an open FRW model} 
\setcounter{equation}{0}

We now turn to the case in which a plane gravitational wave propagates in an
open FRW universe with a stiff perfect fluid. In this case, it is convenient
to adopt the coordinates $\eta,\mu,x,y$ as defined in case~2a of Section~5.
In the background region, the metric functions are given by (\ref{5.6}) and
the fluid by (\ref{5.12}). Since $e^{-U}=\sinh2\eta\,e^{-2\mu}$ in this case,
we can put 
 \begin{equation}
\label{9.1}
f={\textstyle{1\over2}}\left(e^{2(\eta-\mu)}-1\right), \qquad 
g={\textstyle{1\over2}}\left(1-e^{-2(\eta+\mu)}\right). 
 \end{equation}
 We can now consider a gravitational wave with a plane wavefront given by
$\eta=\mu$ as described in Section~3, with the region $\eta<\mu$ being part
of the exact open FRW model into which it propagates.

Since $V=0$ in the background region in this case, the gravitational wave can
be introduced by putting 
 \begin{equation}
 \label{9.2} 
 V(\eta,\mu) =\Theta(\eta-\mu) \int_{1/2}^\infty \phi(k)
\sinh^k2\eta\,e^{-2k\mu}  H_k\left({\textstyle
{e^{2\mu}-\cosh2\eta\over\sinh2\eta} }\right) dk, 
 \end{equation} 
 where $H_k$ is most conveniently given by (\ref{6.12}) and its extension. 
After applying again the method described in \cite{AlGr95}, the complete
integral for $M$ is found to be 
 \begin{equation}
 \label{9.3} 
 M(\eta,\mu) =-\log\sinh2\eta +\tilde M\,\Theta(\eta-\mu), 
 \end{equation} 
 where 
 \begin{eqnarray}
 \label{9.4} 
 &&\hskip-12pt \tilde M =-{\textstyle{1\over2}} \int_1^\infty
\sinh^n2\eta\,e^{-2n\mu} {1\over n}dn \int_{1/2}^{n-1/2} \phi(k)\phi(n-k)
\nonumber\\
 &&\qquad\qquad\qquad 
\left[k(n-k)H_kH_{n-k}
-\left({\textstyle{1-2\cosh2\eta\,e^{2\mu}+e^{4\mu}\over\sinh^22\eta}}\right)
H_{k-1}H_{n-k-1}\right] dk, 
 \end{eqnarray} 
 in which the argument of $H_k$ is $(e^{2\mu}-\cosh2\eta)/\sinh2\eta$.

As previously, we can determine the behaviour of the gravitational wave near
the wavefront where $\eta-\mu$ is small by considering the lowest non-zero
term in $k$ in (\ref{9.2}). Taking this lowest term in the form
 \begin{equation}
 \label{9.5} 
 V(\eta,\mu) ={a_k\over c_k} \sinh^k2\eta\,e^{-2k\mu}  H_k\left({\textstyle
{e^{2\mu}-\cosh2\eta\over\sinh2\eta} }\right)
\Theta(\eta-z), 
 \end{equation} 
 it can be seen that, near the wavefront, the non-zero Weyl tensor components
behave as 
 \begin{eqnarray}
 \label{9.6} 
 &&\hskip-10pt \Psi_4 \sim-{a_k\,(k+{1\over2})\over2^{1/2}(e^{4\eta}-1)^{3/2}}
(\eta-\mu)^{k-1/2} \,\delta(\eta-\mu)
-{a_k\,(k^2-{1\over4})\over2^{1/2}(e^{4\eta}-1)^{3/2}}\, (\eta-\mu)^{k-3/2}
\,\Theta(\eta-\mu), \nonumber\\
 &&\hskip-10pt \Psi_2 \sim -{{a_k}^2(k+{1\over2})\over4\sinh^32\eta}
(\eta-\mu)^{2k} \,\Theta(\eta-\mu), \\
 &&\hskip-10pt \Psi_0 \sim-{3a_k\,e^{5\eta}\over2\,\sinh^{7/2}2\eta}
(\eta-\mu)^{k+1/2} \,\Theta(\eta-\mu). \nonumber
 \end{eqnarray}
 It can thus be seen that an impulsive gravitational wave is again included
here if the lowest term in the expansion for $V$ has $k={1\over2}$. It has a
step (or shock) gravitational wave if the lowest term has $k={3\over2}$.
Generally, the Weyl tensor is $C^n$ across the wavefront if
$k_{\rm min}=n+{5\over2}$.

{}From the above results, it can be seen that these waves in open FRW
universes are qualitatively similar to those in spatially flat FRW universes.
They both include impulsive, shock or smooth fronted wavefronts. As discussed
in Section~3, the geometrical properties of these plane wavefronts in open
FRW universes are described in figure~1 and equation~(\ref{3.7}).

A discussion of these waves and their wavefonts, as well as the possible
collision of such waves, was given in \cite{BiGr94}. However, the complete
solution for $M$ and an analysis of the character of the Weyl tensor
components near the wavefront has been included here for the first time.

\section{Cylindrical waves in an open FRW model} 
\setcounter{equation}{0}

We now consider a gravitational wave which propagates into a stiff fluid open
FRW universe along the null cylindrical wavefronts described in figure~2 and
Eq. (\ref{3.10}). In this case, it is convenient to adopt the coordinates
$\eta,\rho,\phi,z$ as defined in case~2b of Section~5. In the background
region, the metric functions are given by (\ref{5.7}) and the fluid by
(\ref{5.12}). Since $e^{-U}=\sinh2\eta\sinh2\rho$ in this case, we can put 
 \begin{equation}
\label{10.1}
f=-\sinh^2(\eta-\rho), \qquad g=\sinh^2(\eta+\rho). 
 \end{equation}

We can now consider a gravitational wave with a cylindrical wavefront
$\eta=\rho$ as described in Section~3, with the region $\eta<\rho$ being part
of the exact open FRW universe into which the wave propagates. This can be
achieved by putting 
 \begin{equation}
 \label{10.2} 
 V(\eta,\rho) =\log\tanh\rho +\tilde V(\eta,\rho)\Theta(\eta-\rho), 
 \end{equation} 
 where
 \begin{equation}
 \label{10.3} 
 \tilde V(\eta,\rho) =\int_{1/2}^\infty \phi(k)
\sinh^k2\eta\sinh^k2\rho\, H_k\left({\textstyle
{\cosh2\eta\cosh2\rho-1\over\sinh2\eta\sinh2\rho} }\right) dk, 
 \end{equation} 
 in which $H_k$ is now most conveniently given by (\ref{6.11}). In this case
the integration of the subsidiary equations (\ref{6.7}) is more complicated.
However, the complete integral for $M$ can still be obtained and expressed in
the form 
 \begin{equation}
 \label{10.4} 
 M(\eta,\rho) =-\log\sinh2\eta +\tilde M(\eta,\rho)\,\Theta(\eta-\rho), 
 \end{equation} 
 where 
 \begin{eqnarray}
 \label{10.5} 
 &&\tilde M =-{\textstyle{1\over2}} \int_{1/2}^\infty \phi(k) \left\{
2k\,A_{k-1}(\tau,\zeta)H_k +\left[(\zeta^2-1)^{1/2}{1\over k}\,\tau^k
+(\zeta^2-1)A_k(\tau,\zeta)\right] H_{k-1} \right\} dk \nonumber\\
 &&\qquad\qquad -{\textstyle{1\over2}} \int_1^\infty {1\over n}\,\tau^n dn
\int_{1/2}^{n-1/2} \phi(k)\phi(n-k) \nonumber\\ 
 &&\qquad\qquad\qquad\qquad 
\left[k(n-k)H_kH_{n-k} -(\zeta^2-1) H_{k-1}H_{n-k-1}\right] dk, 
 \end{eqnarray} 
 in which
 \begin{equation}
A_k(\tau,\zeta)=\int_0^\tau {t^k\over[(2+\zeta t)^2-t^2]^{1/2}}\, dt,
 \end{equation}
$H_k=H_k(\zeta)$ are as defined above, and $\tau$ and $\zeta$ are functions
of $\eta$ and $\rho$ given by 
 \begin{equation}
\tau=\sinh2\eta\sinh2\rho, \qquad
\zeta={\cosh2\eta\cosh2\rho-1\over\sinh2\eta\sinh2\rho}. 
 \end{equation}

It can be seen that these solutions are singular at the initial big bang at
which $\eta=0$, and that there is an additional curvature singularity on the
axis of symmetry $\rho=0$. The singular axis can be considered to represent
both the source of the cylindrical gravitational wave and the sink of the
backscattered radiation, as in the spatially flat case described in
Section~8.

It is also possible to analyse the character of the wave along the wavefront
using the same method as in previous sections. Accordingly, we include only
the lowest non-zero term in $k$ in (\ref{10.2}) in the form
 \begin{equation}
 \label{10.8} 
 \tilde V(\eta,\rho) ={a_k\over c_k} \sinh^k2\eta\sinh^k2\rho\,
H_k\left({\textstyle {\cosh2\eta\cosh2\rho-1\over\sinh2\eta\sinh2\rho}
}\right). 
 \end{equation} 
 With this it can be shown that, near the wavefront, the non-zero Weyl
tensor components behave as 
 \begin{eqnarray}
 \label{10.9} 
 &&\Psi_4 \sim-{a_k\,(2k+1)(4k+1)\over2\,\sinh^22\eta}\,
(\eta-\rho)^{2k-1} \,\Theta(\eta-\rho), \nonumber\\
 &&\Psi_2 \sim {a_k\,(2k+1)\over2\,\sinh^32\eta}\, (\eta-\rho)^{2k}
\,\Theta(\eta-\rho), \\
 &&\Psi_0 \sim-{3\,a_k\over2\,\sinh^42\eta}\, (\eta-\rho)^{2k+1}
\,\Theta(\eta-\rho). \nonumber
 \end{eqnarray}
 These expressions can be seen to be very similar to those for cylindrical
waves in the spatially flat FRW universe as described in Section~8 (see
(\ref{8.7}) in particular). As there, it can be seen that these solutions
include a step (or shock) gravitational wave if the lowest term in the
expansion for $V$ has $k={1\over2}$. Generally, the Weyl tensor is $C^n$
across the wavefront if $k_{\rm min}={1\over2}n+1$.

The possibility of obtaining explicit solutions of this type was mentioned in
\cite{AlGr95} but the solutions are described in detail here for the first
time. It is also possible to consider the collision of opposing outgoing and
incoming cylindrical gravitational waves just as in the spatially flat case.
However, although the qualitative features of such a situation can be
determined, a complete integral for $M$ in the interaction region cannot be
obtained in this case using the above method.

\section{Toroidal waves in a closed FRW model} 
\setcounter{equation}{0}

We now consider waves in a closed FRW universe. In this case, no
plane-fronted waves can exist. However, as described in Section~4, waves with
toroidal surfaces can be constructed. These are analogous with the
cylindrical-fronted waves in an open universe considered in the previous
section. Mathematically, they can be obtained by replacing the hyperbolic
functions by the corresponding trigonometric functions. Physically, the
cylindrical wavefront becomes closed to form a toroidal wavefront.

In this case, it is convenient to adopt the coordinates
$\eta,\zeta,\delta,\sigma$ as defined in case~3 of Section~5. Starting from
the background (\ref{5.8}), we may consider the family of null hypersurfaces
$\eta-\zeta=$~const. At any fixed time $\eta=$~const., the wave surface is a
toroid $\zeta=$~const. However, as explained in Section~4, two axes now exist
--- at $\zeta=0$ and at $\zeta=\pi/2$. We may consider a wave starting at the
initial big bang singularity ($\eta=0$) at the circular axis $\zeta=0$, and
propagating along the null hypersurface $\eta=\zeta$ into the FRW closed
universe. At any fixed time $\eta$, the wavefront will be a toroidal
2-surface as illustrated in figure~4. As the universe expands, the front of
the wave propagates towards the (circular) axis $\zeta=\pi/2$.

In the background region, the metric functions are given by (\ref{5.8}) and
the fluid by (\ref{5.13}). Since $e^{-U}=\sin2\eta\sin2\zeta$ in this case,
we can put 
 \begin{equation}
\label{11.1}
f=-\sin^2(\eta-\zeta), \qquad g=\sin^2(\eta+\zeta). 
 \end{equation}
 The gravitational wave with the toroidal wavefront $\eta=\zeta$ is
determined by putting
 \begin{equation}
 \label{11.2} 
 V(\eta,\zeta) =\log\tan\zeta +\tilde V(\eta,\zeta)\Theta(\eta-\zeta), 
 \end{equation} 
 where
 \begin{equation}
 \label{11.3} 
 \tilde V(\eta,\zeta) =\int_{1/2}^\infty \phi(k)
\sin^k2\eta\sin^k2\zeta\, H_k\left({\textstyle
{1-\cos2\eta\cos2\zeta\over\sin2\eta\sin2\zeta} }\right) dk, 
 \end{equation} 
 in which $H_k$ is given by (\ref{6.11}). As above, it can be shown that the
complete integral for $M$ can be expressed in the form 
 \begin{equation}
 \label{11.4} 
 M(\eta,\zeta) =-\log\sin2\eta +\tilde M(\eta,\zeta)\,\Theta(\eta-\zeta), 
 \end{equation} 
 where 
 \begin{eqnarray}
 \label{11.5} 
 &&\tilde M =-{\textstyle{1\over2}} \int_{1/2}^\infty \phi(k) \left\{
2k\,B_{k-1}(\tau,\xi)H_k +\left[(\xi^2-1)^{1/2}{1\over k}\,\tau^k
-(\xi^2-1)B_k(\tau,\xi)\right] H_{k-1} \right\} dk \nonumber\\
 &&\qquad\qquad -{\textstyle{1\over2}} \int_1^\infty {1\over n}\,\tau^n dn
\int_{1/2}^{n-1/2} \phi(k)\phi(n-k) \nonumber\\ 
 &&\qquad\qquad\qquad\qquad 
\left[k(n-k)H_kH_{n-k} -(\xi^2-1) H_{k-1}H_{n-k-1}\right] dk, 
 \end{eqnarray} 
 in which
 \begin{equation}
B_k(\tau,\xi)=\int_0^\tau {t^k\over[(2-\xi t)^2-t^2]^{1/2}}\, dt,
 \end{equation}
 and $\tau$ and $\xi$ are functions of $\eta$ and $\zeta$ given by 
 \begin{equation}
\tau=\sin2\eta\sin2\zeta, \qquad
\xi={1-\cos2\eta\cos2\zeta\over\sin2\eta\sin2\zeta}, 
 \end{equation}
 and to avoid confusion with the coordinate $\zeta$, in this case, we have
relabelled the argument of $H_k$ putting $H_k=H_k(\xi)$.

The character of the wave along the wavefront can also be analysed as in
previous sections. Accordingly, we consider the lowest non-zero term in $k$
in (\ref{11.2}) in the form
 \begin{equation}
 \label{11.8} 
 \tilde V(\eta,\zeta) ={a_k\over c_k} \sin^k2\eta\sin^k2\zeta\,
H_k\left({\textstyle {1-\cos2\eta\cos2\zeta\over\sin2\eta\sin2\zeta}
}\right). 
 \end{equation} 
 With this it can be shown that, near the wavefront, the non-zero Weyl
tensor components behave as 
 \begin{eqnarray}
 \label{11.9} 
 &&\Psi_4 \sim-{a_k\,(2k+1)(4k+1)\over2\,\sin^22\eta}\,
(\eta-\zeta)^{2k-1} \,\Theta(\eta-\zeta), \nonumber\\
 &&\Psi_2 \sim {a_k\,(2k+1)\over2\,\sin^32\eta}\, (\eta-\zeta)^{2k}
\,\Theta(\eta-\zeta), \\
 &&\Psi_0 \sim-{3\,a_k\over2\,\sin^42\eta}\, (\eta-\zeta)^{2k+1}
\,\Theta(\eta-\zeta). \nonumber
 \end{eqnarray}
 These expressions are obvious modifications of those for cylindrical waves
in spatially flat and open FRW universes as given by  (\ref{8.7}) and
(\ref{10.9}). They include a step (or shock) gravitational wave if the lowest
term in the expansion for $V$ has $k={1\over2}$. Generally, the Weyl tensor
is $C^n$ across the wavefront if $k_{\rm min}={1\over2}n+1$.

It can be seen that these solutions are singular at the initial big bang at
which $\eta=0$, and at the final big crunch at which $\eta=\pi/2$. There is
also an additional curvature singularity on the axis of symmetry $\zeta=0$,
which can be considered to represent both the source of the toroidal
gravitational wave and the sink of the backscattered radiation.

As noticed in Section~5, from the big bang until the big crunch, a photon in
a closed universe with a stiff fluid succeeds in travelling only one quarter
of the distance around the whole universe. Similarly, an exact shock wave
starting at the big bang on the ``polar axis'' $\zeta=0$ will reconverge onto
the ``equatorial'' axis $\zeta=\pi/2$ at the big crunch. The space-time
diagram of such a situation is given in figure~5.

It is also possible to consider the collision of two shock waves propagating
into the closed FRW background with a stiff fluid. At the big bang, one wave
starts at the axis $\zeta=0$ and the other at $\zeta=\pi/2$. As the universe
expands, the waves approach each other and collide just at the time of
maximum expansion. One may also consider other initial conditions with the
waves being some distance from the axes at time $\eta=0$. Such waves will
collide while the universe is still expanding. However, although the
qualitative features of such a situation can be determined a complete
integral for $M$ cannot be obtained in this case. Locally, such waves were
considered in \cite{FeGr94}. (The coordinates $z,x,y$ used in \cite{FeGr94}
are related to those above by $\zeta=z$, $\delta=2^{1/2}x$,
$\sigma=2^{1/2}y$.) The toroidal character of the waves was, however, not
noticed and the discussion of the singularities was incorrect.

Solutions of this type are closed universes with two global, commuting and
hypersurface orthogonal Killing vectors. They may therefore be considered as
Gowdy-type space-times \cite{Gow74}, \cite{Gow75}. It can readily be verified
that here the regularity conditions across the null hypersurface $\eta=\zeta$
are satisfied. Such conditions were studied carefully by Gowdy himself in the
vacuum case, and by Kitchingham \cite{Kit84} (see also references therein) in
the case of stiff fluid cosmologies.

\section{ Concluding remarks} 
\setcounter{equation}{0}

Exact gravitational waves with two isometries have been constructed which can
be considered as exact perturbations of FRW cosmological models with a stiff
fluid. The waves have distinct wavefronts and propagate into any of the three
possible FRW backgrounds. Although much work has been done on considering
various types of waves in the context of cosmological models (c.f. reviews
\cite{CCM81}--\cite{Kra95}), waves with distinct wavefronts propagating into
the standard FRW backgrounds have only recently been considered
(\cite{Gri93a}--\cite{AlGr95}).

In the spatially flat and open universes, the wavefronts can be either plane
or cylindrical. In the closed models, they are toroidal. One may consider
toroidal waves propagating in other models as well, provided that the spatial
sections are compactified by making appropriate identifications. However,
such identifications result in anisotropic background spaces. Here we have
assumed the standard backgrounds with natural topologies admitting globally
the six-parameter group of isometries.

We have obtained exact solutions for plane-fronted gravitational waves
propagating in a spatially flat or open FRW universe with a stiff fluid.
These include impulsive waves, shock waves, or waves with Weyl tensor
components of any arbitrary finite degree of smoothness. We have also
presented exact solutions for cylindrical- or toroidal-fronted gravitational
waves propagating in a spatially flat, open or closed FRW stiff fluid
universe. These do not include impulsive waves, but may include shock waves
or waves of any arbitrary finite degree of smoothness.

Solitonic perturbations of FRW models with cylindrical symmetry, without
wavefronts, have commonly been treated in the literature (see e.g. 
\cite{Bel79} and \cite{Kit84}, and references therein). The
possibility of constructing plane-fronted waves in the open FRW universes
appears to have been recognised only recently \cite{BiGr94}. On the other
hand, it is well known that the $G_2$ cosmological models, that are often
interpreted as inhomogeneous generalisations of the FRW models, can sometimes
also be used to describe cylindrical perturbations simply by reinterpreting
the relevant coordinates. It can be seen from the above examples that, not
only is the topological character of the wavefronts changed by this
procedure, but the character of the wave may be altered as well.

No singularities, except those representing the big bang or big crunch, arise
in the closed vacuum models considered by Gowdy \cite{Gow74}, \cite{Gow75}
or in the soliton-type waves studied by Belinskii \cite{Bel79}, Kitchingham
\cite{Kit84} and others (see \cite{CCM81}, \cite{Ver93}, \cite{Kra95} for
references) in cases of both open and closed FRW stiff fluid models. Their
explicit solutions, although without wavefronts, clearly correspond to our
cylindrical waves, or to toroidal waves in the closed case. However, in the
spatially flat and open FRW models, they could also be adapted to infinite
plane perturbations.

With a clear geometrical picture of the wavefronts available, it may be of
interest to consider also other types of shock waves with two isometries
propagating into these FRW backgrounds. These may be just test radiative
fields (not necessarily gravitational), or exact (large) perturbations. For
example, one could discuss fluid shocks in all types of FRW models, as
Tabensky and Taub \cite{TaTa73} considered shock acoustic waves in the
spatially flat universe in their pioneering work.

The advantage of studying shock gravitational waves with distinct wavefronts
is not only in having a clear picture of the wave surfaces. One can also
discuss the head-on collision of such waves. In the case of the spatially
flat and open universes, the collision of waves with plane surfaces was
analysed in detail in \cite{Gri93b} and \cite{BiGr94}. The presence of the
gravitational waves slows down the rate of expansion, but future spacelike
singularities do not occur for interacting waves in these expanding universes
as they do in the vacuum case with a Minkowski background. This result is
qualitatively similar to those of Centrella \cite{Cen80} and Centrella and
Matzner \cite{CeMa82} which study the collision of waves in an expanding
vacuum Kasner background.

Similarly, one can consider the collision of ``outgoing'' and ``incoming''
waves with cylindrical wavefronts in the spatially flat and open universes,
and toroidal wavefronts in the closed models. No future spacelike
singularities seem to arise again. However, it appears that making the
wave surfaces finite along one direction in the cylindrical case and fully
finite in the toroidal case gives rise to timelike singularities representing
the histories of the corresponding axes. The ``source'' at the axis appears
to be needed to support the propagation of such waves.

\section*{Appendix}
\setcounter{section}{0}
\renewcommand{\thesection}{\Alph{section}}
\renewcommand{\theequation}{\thesection{A.}\arabic{equation}}

We here consider the geometry of possible wave surfaces in a closed FRW
universe in greater detail. We will show that a typical wave surface is a
2-torus in the 3-sphere described by (\ref{4.3}) and (\ref{4.4}). First
consider a circle {\it S}, as illustrated in figure~6, in the $W,Z$-plane
given by 
 \begin{equation}
 \label{A.1} 
 {\it S}: \qquad W^2+Z^2=\kappa^2R_0^2, \qquad X=Y=0, 
 \end{equation}
 where $\kappa\in[0,1]$ is a constant. The circle
{\it S} is generally inside the 3-sphere. Next, consider a 2-plane given by
 \begin{equation}
 \label{A.2} 
 (X,Y)\in{\hbox{$I\kern-3pt R$}}^2, \quad W,Z~{\rm fixed}, \quad
W^2+Z^2=\kappa^2R_0^2. 
 \end{equation}
 Such a plane for any fixed $W$ and $Z$ touches the circle {\it S} at just
one point, and it intersects the 3-sphere in the circle
 \begin{equation}
 \label{A.3} 
 X^2+Y^2=(1-\kappa^2)R_0^2, 
 \end{equation}
 with $W,Z$ fixed and satisfying $W^2+Z^2=\kappa^2R_0^2$. When $W=\kappa
R_0$ and $Z=0$, this is the circle {\it C} illustrated in figure~7. In
figure~6 only two points $a$ and $b$ of this circle can be seen because one
dimension has been suppressed. Now by moving the point $(W,X)$ along the
circle {\it S} and considering the corresponding family of 2-planes
(\ref{A.2}), we find that the planes intersect the 3-sphere in the 2-torus
$T^2=S^1\otimes S^1$ given by
 \begin{equation}
 \label{A.4} 
 W^2+Z^2=\kappa^2R_0^2, \qquad X^2+Y^2=(1-\kappa^2)R_0^2. 
 \end{equation}

In figure~6, the slice $Y=0$ through the 2-torus is seen as two circles, $A$
and $B$, given by (\ref{A.4}) with $Y=0$. These circles represent the
intersection of the 2-sphere $W^2+Z^2+X^2=R_0^2$ through the 2-torus
(\ref{A.4}). In a universe with large $R_0$, the circles will locally
approximate to straight lines and, if $\kappa$ is close to 1, the surfaces
will appear to be cylindrical.

Suppressing the coordinate $Z$, the 2-torus is represented by the circles
$X^2+Y^2=(1-\kappa^2)R_0^2$, $W=\pm\kappa R_0$, denoted by $C$ and $D$ in
figure~7. These two circles represent the intersection of the 2-sphere
$X^2+Y^2+W^2=R_0^2$ through the 2-torus (\ref{A.4}).

It is seen from (\ref{A.4}) that the 2-torus is specified by the choice of
the parameter $\kappa$ which may conveniently be written as
$\kappa=\cos\zeta$, with $0\le\zeta\le\pi/2$. With the form of (\ref{A.4}),
the natural coordinates on the 2-torus are the angles, say  $\sigma$ and
$\delta$, parametrizing the two circles of radii $R_0\cos\zeta$ and
$R_0\sin\zeta$. This leads to the alternative parametrization of the whole
3-sphere given by
 \begin{eqnarray}
 \label{A.5} 
 W &=& R_0\cos\zeta\cos\sigma, \nonumber\\
Z &=& R_0\cos\zeta\sin\sigma, \nonumber\\
X &=& R_0\sin\zeta\cos\delta, \nonumber\\
Y &=& R_0\sin\zeta\sin\delta, 
\end{eqnarray} 
 where $\zeta\in[0,\pi/2]$, $\sigma\in[0,2\pi]$, $\delta\in[0,2\pi]$. In
figure~6, the two families of circles $A$ and $B$, each of radius
$R_0\cos\zeta$, have $\delta=0$ and $\delta=\pi$ respectively, and are
parametrized by $0\le\sigma\le2\pi$. Similarly, the families of circles $C$
and $D$, illustrated in figure~7, have radius $R_0\sin\zeta$, with $\sigma=0$
and $\sigma=\pi$ respectively, and are parametrized by $0\le\delta\le2\pi$.
Thus, this is a natural  parametrization for the description of toroidal
2-surfaces inside the 3-sphere. Notice that degenerate points occur in
both figures either when $\zeta=0$, or when $\zeta=\pi/2$. These are
coordinate singularities in which the norms of the Killing vectors
$\partial_\delta$ or $\partial_\sigma$ vanish. It is easily verified that the
3-sphere is covered once with the angular variables $\zeta$, $\sigma$,
$\delta$ taking values in the given intervals. On the 3-sphere, these angular
variables are related to the familiar Euler angular coordinates $\theta$,
$\phi$, $\psi$ (which should not be confused with the angles used in
(\ref{4.1}) etc.) by $\theta=2\zeta$, $\phi=\sigma-\delta$,
$\psi=\sigma+\delta$. The Euler angles are also often used since the 3-sphere
has the same topology as the group $SU(2)$, and the generators of the
symmetries can be suitably expressed in terms of these angles (see e.g.
\cite{Gow74} or \cite{BeDr70}).

It is possible to consider the 2-torus (\ref{A.4}) as dividing the 3-sphere
into two solid tori. Conversely, $S^3$ can be constructed by identifying the
boundaries of two solid tori. This appears to have been first explained in
relativity literature by Misner \cite{Mis63} in his analysis of Taub--NUT
space, using an inversion. We will now demonstrate this, using the
parametrization (\ref{A.1}), thus making apparent the character of the
coordinates $\zeta$, $\sigma$ and $\delta$. (See also Appendix B of
\cite{Gow74}.)

We first cut the 3-sphere (\ref{4.3}) into two along the hypersurface $Z=0$,
or $\sigma=0$ and $\sigma=\pi$, to form two {\it solid} 2-spheres. (This is
similar to cutting an ordinary 2-sphere along an equatorial plane to form two
hemispheres which can each be mapped, by stereographic projection, onto a
disc.) This is illustrated in figure~8, in which the solid 2-sphere on the
left corresponds to the part of the 3-sphere in which $Z\ge0$, that on the
right to $Z\le0$. The two spheres are reconnected by identifying
corresponding points on their surfaces.

The full curves inside the two spheres indicated in figure~8 represent the
section of the sphere in which $\delta=0$ and $\delta=\pi$ for some fixed
value of $\zeta$. They can be seen to arise from the stereographic projection
of the hemispheres with $Z\ge0$ and $Z\le0$ in figure~6 onto the $(W,X)$
plane, the centre of projection being the points $W=X=0$, $Z=-R_0$ and
$W=X=0$, $Z=+R_0$. Introducing coordinates $\tilde W$, $\tilde X$ for the
projected points of the hemisphere with, say $Z\ge0$, $\tilde
W=WR_0/(R_0+Z)$, $\tilde X=XR_0/(R_0+Z)$, and substituting for $W$, $X$ and
$Z$ from (\ref{A.5}) with $\delta=0$, we obtain $\tilde
W^2+(\tilde X-R_0/\sin\zeta)^2=R_0^2\cot\zeta$. For fixed values of $\zeta$,
these are segments of circles, such as $A$ in figure~8, with centres on
$\tilde X>0$, bounded by $Z=0$. Along these segments $\sigma$ is changing and
$\delta=0$. As $\zeta\to\pi/2$, the segments degenerate to the ``East pole''
on the equator. It can thus be seen that the circles $A$ and $B$ in figure~6
are  projected into the corresponding circles in figure~8.

The surfaces of both solid 2-spheres, defined by $Z=0$, must be identified in
order to reconstruct the original 3-sphere. In figure~8, such an
identification is illustrated in terms of the coordinates $\zeta$, $\sigma$
and $\delta$. Surfaces inside each sphere on which $\zeta$ is constant, as
indicated in figure~8, are joined on the surface of each sphere, dividing the
3-sphere into two solid tori. One torus has the closed axis given by
$\zeta=0$ and is formed by the axes of the two spheres illustrated in
figure~8. On this axis $\sigma$ changes from 0 to $\pi$ in the left sphere
and continues from $\pi$ up to $2\pi$ in the right sphere. Along this axis,
the coordinate $\delta$ is degenerate and the norm of the Killing vector
$\partial_\delta$ vanishes. The second torus is evident on identifying points
near the equators of both spheres. The equators form the second closed axis.
It is given by $\zeta=\pi/2$ and is parametrized by $0\le\delta\le2\pi$, while
the coordinate $\sigma$ is degenerate and the norm of the Killing vector
$\partial_\sigma$ vanishes. These two axes, $\zeta=0$ and $\zeta=\pi/2$, are
crucial in dealing with toroidal waves in closed FRW universes.

\section*{acknowledgements}

JB was supported by an award from the Royal Society and, in part, by the
grant GACR-202/96/0206 of the Czech Republic and the U.S.-Czech Science and
Technology grant no. 92067.

\vfill\eject

\begin{figure}
\caption{ The embedding diagram of the section $Y=0$ (or $\phi=0$ and $\pi$)
through the 3-hyperboloid (3.3) in a 3-dimensional Minkowski space. The
2-planes given by $z=z_0=$~const., $(x,y)\in{\hbox{$I\kern-3pt R$}}^2$ (or
$(X,Y)\in{\hbox{$I\kern-3pt R$}}^2$) are illustrated as parabolae along which
$x,y$ vary. }
\label{fig1}
\end{figure}

\begin{figure}
\caption{ The embedding diagram of the section $Y=0$ (or $\phi=0$ and $\pi$)
through the 3-hyperboloid (3.3) in a 3-dimensional Minkowski space. The
cylindrical 2-surfaces given by $\rho=\rho_0=$~const.,
$y\in(-\infty,+\infty)$, $\phi\in[0,2\pi)$ are illustrated in two hyperbolae
representing two generators of the cylinder along which the coordinate $y$
varies. The axis of the cylinder given by $\rho=0$ is also indicated. }
\label{fig2}
\end{figure}

\begin{figure}
\caption{ A space-time diagram illustrating the collision of cylindrical
gravitational waves in a spatially flat or open FRW universe.  Region~I is
the FRW background in cylindrical coordinates (5.5) or (5.7). Region~II
contains an outgoing cylindrical wave, region~III contains an incoming
cylindrical wave, and region~IV is the interaction region following their
collision. The space-time is only singular on the initial hypersurface
$\eta=0$, and on the axis $\rho=0$. }
\label{fig3}
\end{figure}

\begin{figure}
\caption{ Toroidal waves propagating into a closed FRW background. The figure
shows two sections through the toroid as in figures 6 and 7 in the Appendix.
The unshaded regions represent the exact FRW model, with the wave  region
illustrated schematically by the shaded areas. The singular circular axis
$\zeta=0$ within the wave region is also indicated.  }
\label{fig4}
\end{figure}

\begin{figure}
\caption{ This space-time diagram illustrates the propagation of a
gravitational wave into a closed FRW universe. The coordinates $\delta$ and
$\sigma$ are suppressed. The shaded region contains the gravitational wave.
The null wavefront is toroidal as illustrated in figure~4. It starts at a
line at the initial singularity $\eta=0$ and converges to another line at the
final singularity $\eta=\pi/2$. There is also a curvature singularity on the
initial axis $\zeta=0$. }
\label{fig5}
\end{figure}

\begin{figure}
\caption{ A section $Y=0$ ($\phi=0$ and $\pi$) through the 3-sphere of radius
$R_0$ showing circular cuts $A$ and $B$ through the toroidal surface. The
circles $S$, $A$ and $B$ have radius $\kappa R_0$. }
\label{fig6}
\end{figure}

\begin{figure}
\caption{ A section $Z=0$ ($\theta=\pi/2$) through the 3-sphere of radius
$R_0$ showing circular cuts $C$ and $D$ through the toroidal surface. The
points $a$, $a'$ and $b$, $b'$ are on the circles $A$ and $B$ respectively,
indicated in figure~6. }
\label{fig7}
\end{figure}

\begin{figure}
\caption{ The entire spatial geometry of the closed FRW model at a given time
is represented by two solid spheres. Points on the surface of upper
hemisphere on the left on which $\sigma=0$ must be identified with points on
the surface of the upper hemisphere on the right on which $\sigma=2\pi$.
Points on the surfaces of both lower hemispheres on which $\sigma=\pi$ must
also be identified. Within each sphere, the segments are indicated which
arise from the section $\delta=0$ and $\pi$ through the spheres. They
represent the stereographic projections of the circles $A$ and $B$ in
figure~6 onto the ($W,X$) plane. }
\label{fig8}
\end{figure}

\end{document}